# Phase equilibria in $MnSb_2Te_4$ – $GeSb_2Te_4$ system and magnetic properties of $Mn_{1-x}Ge_xSb_2Te_4$ solid solutions

Fuad Safarov [a,b,c,*], Dunya Babanly [a,b,d], Jerome Robert [c], Elnur Orujlu [b], Elvin Ahmadov [d,e], Martin Bowen [c], Bohdan Kundys [c]

[a] *French-Azerbaijani University (UFAZ), 183 Nizami str., AZ 1000 Baku, Azerbaijan*
[b] *Azerbaijan State Oil and Industry University (ASOIU), 34 Azadliq ave., AZ 1010, Baku, Azerbaijan*
[c] *Université de Strasbourg, CNRS, Institut de Physique et Chimie des Matériaux de Strasbourg, UMR 7504, Strasbourg F-67000, France*
[d] *Institute of Catalysis and Inorganic Chemistry (ICIC), Azerbaijan National Academy of Sciences (ANAS), 113 H. Javid str., AZ 1143, Baku, Azerbaijan*
[e] *Azerbaijan State University of Economics (UNEC), 6 Istiglaliyyat str., AZ 1001 Baku, Azerbaijan*

ABSTRACT

As a sister compound of the antiferromagnetic topological insulator $MnBi_2Te_4$, $MnSb_2Te_4$ is also a candidate for exotic magnetic topological phases. On the other hand, the structurally analogous but nonmagnetic phase-change material $GeSb_2Te_4$ is also known to exhibit nontrivial band topology. Motivated by their shared crystal structure, the similar ionic radii of $Mn^{2+}$ and $Ge^{2+}$, and the opportunity to explore the interplay between magnetism and topology, here we investigate the effects of Ge substitution at Mn sites in $MnSb_2Te_4$. The $Mn_{1-x}Ge_xSb_2Te_4$ solid solutions were synthesized via high-temperature solid-state reaction and characterized for composition, structure, phase behavior, and magnetism using SEM-EDS, PXRD, DTA, and SQUID. Ge substitution was successful across the full composition range, producing homogeneous, single-phase samples that melt via peritectic reactions, as confirmed by the $MnSb_2Te_4$ – $GeSb_2Te_4$ phase diagram. Ge substitution strengthens the sample's paramagnetism, but with ferrimagnetic ordering up to $x = 0.75$, with both effective moment and saturation magnetization decreasing with increasing Ge content. Two distinct magnetic transitions – high-temperature paramagnetic to ferrimagnetic and low-temperature ferrimagnetic to ferromagnetic – were identified, with a dome-like shape dependence of the low-temperature magnetic transition on Ge substitution. A negative magnetization was observed in the pristine $MnSb_2Te_4$ and $x = 0.12$ substituted samples, while two distinct spin-flop transitions appeared in the samples with $x = 0.32$ and $x = 0.55$ Ge substitutions as a result of competing magnetic orderings. These findings facilitate future selective single-crystal growth of homogeneous, magnetic phases, paving the way for magneto-transport and topological surface states investigations.

## 1. Introduction

Topological insulators are a class of quantum materials with gapless surface states that are protected by time reversal symmetry and insulating bulk states [1,2]. Their surface states are robust against nonmagnetic disorder and characterized by spin-momentum locking, indicating exciting opportunities for applications in spintronics and low-power electronics [3,4]. When magnetism is introduced into nonmagnetic topological insulators, time-reversal symmetry is broken, leading to exotic quantum phenomena such as the quantum anomalous Hall effect (QAHE) and the axion insulator (AXI) state [5–7]. Two traditional methods of introducing magnetism into TIs are doping with magnetic atoms [5,8,9] and creating TI/magnet heterostructures [9–11]; however, the resulting magnetic systems are far from being satisfactory as doping tends to cause inhomogeneities in the doped bulk, and heterostructures often suffer from problems related to interfacial band bending [10].

In contrast, intrinsic magnetic topological insulators offer a cleaner platform without the complications of disorder or complex interface physics associated with doped or proximity-coupled systems [5,12–16]. The realization of intrinsic magnetic topological insulators not only advances our understanding of quantum materials but also holds promise for next-generation spintronic and topological quantum devices due to their structural simplicity and intrinsic magnetic order. The discovery of the first intrinsic antiferromagnetic topological insulator, $MnBi_2Te_4$, was a major breakthrough in the field of magnetic topological insulators [17]. Both quantum anomalous Hall effect (QAHE) and axion insulator (AXI) states were observed in thin flakes of $MnBi_2Te_4$ under few layer thicknesses and external magnetic field conditions [18,19], highlighting its value as a promising avenue for studying the interplay between magnetism and topology.

* Corresponding author at: French-Azerbaijani University (UFAZ), 183 Nizami str., AZ 1000 Baku, Azerbaijan.
*E-mail address:* fuad.safarov@ipcms.unistra.fr (F. Safarov).

The Mn-rich $MnSb_2Te_4$ was reported to be a ferromagnetic topological insulator in thin-film form [20]. However, neither the quantum anomalous Hall effect nor the axion insulator state has been observed to date, and similar properties have not yet been confirmed in single crystals. Like its sister compound $MnBi_2Te_4$, $MnSb_2Te_4$ had been predicted to exhibit A-type antiferromagnetic (AFM) order, in which Mn magnetic moments in the Mn layer within a septuple layer couple ferromagnetically and those in the neighboring septuple layers couple antiferromagnetically [21]. Though the experimental samples of $MnBi_2Te_4$ exhibit A-type AFM order [17], those of $MnSb_2Te_4$ had been found to be ferromagnetic (FM) due to Mn-Sb cation intermixing, in which Mn-Sb antisite defects play an important role [22]. The Mn-Sb antisite defects have been shown to substantially modify magnetic interactions in $MnSb_2Te_4$ with first-principles calculations and defect modeling, indicating that Mn-Sb occupation promotes ferromagnetic interlayer coupling and drives a transition from antiferromagnetic to ferromagnetic ordering as defect concentration increases [23,24]. Thus, tuning Mn-Sb disorder may provide an effective control parameter to engineer the magnetic gap and enhance the topological insulator properties. With this in mind, we use a nonmagnetic $GeSb_2Te_4$, which is is structurally identical to both $MnSb_2Te_4$ and $MnBi_2Te_4$ magnetic compounds. It is a phase change material reported to manifest nontrivial band topology [25]. Motivated by identical crystal structures of $MnSb_2Te_4$ and $GeSb_2Te_4$, the similar ionic sizes of $Mn^{2+}$ and $Ge^{2+}$, and the potential for exploring interplay between magnetism and topology, in this work, we studied whether Ge substitution at Mn sites in $MnSb_2Te_4$ promotes formation of a homogenous solid solution, and if so up to what solubility limit. We then studied the phase equilibria and magnetic properties.

## 2. Experimental details

### 2.1. Synthesis

Polycrystalline $Mn_{1-x}Ge_xSb_2Te_4$ (MGST-124) quaternary samples were synthesized via a high-temperature solid-state reaction from pre-synthesized and homogenized MnTe, GeTe, and $Sb_2Te_3$ binary compounds. Mn, Ge, Sb, and Te elements with high purity (≥99.999% by weight) in the form of pieces were used as raw materials for synthesis of the binary compounds. Manganese pieces were soaked in nitric acid for 10 s to eliminate metal oxide impurities from their surface, thoroughly rinsed with ethanol, dried, and used right afterwards. To avoid reaction between manganese and quartz at high temperatures during synthesis of manganese telluride, Mn and Te pieces were added into a glassy carbon crucible [26,27], which was enclosed in a quartz tube. Stoichiometric amounts of the elements were accurately weighted using an analytical balance with an accuracy of $\pm 5\cdot 10^{-5}$ g and added into quartz tubes, which were then de-aired and sealed under a residual pressure of $10^{-3}$ Pa. The synthesis was carried out in a commercial muffle furnace by Thermo Scientific, equipped with a PID temperature controller to accurately monitor and control temperature inside the furnace. The temperature was increased to the target value at a ramp rate of 7.5 K/min and maintained at that temperature for 3 h. The target value of the temperature for each binary compound was chosen to be 50 K above its melting temperature, which was obtained from the T-x phase diagram of the corresponding system [28–30]. At the end of the synthesis, the furnace was put to a standby mode and the binaries were allowed to cool to room temperature. Subsequently, they were taken out from the furnace and annealed at a temperature of 723 K for one week in order to achieve a full equilibrium state. After confirming melting temperatures via differential thermal analysis (DTA) and verifying phase purity by comparing powder X-ray diffraction (PXRD) patterns with reference patterns from COD and PDF-4+ databases, the obtained binary compounds were used for synthesis of the MGST-124 samples.

To prepare the $Mn_{1-x}Ge_xSb_2Te_4$ samples with nominal Ge substitutions of $x = 0$, 0.1, 0.3, 0.5, 0.7, 0.9, and 1, stoichiometric amounts of the obtained MnTe, GeTe, and $Sb_2Te_3$ compounds were accurately weighed using an analytical balance and added to quartz ampoules, and were then degassed and sealed under a residual pressure of $10^{-3}$ Pa. For the synthesis via co-melting, the ampoules were heated to 973 K and held at this temperature for 3 h. The temperature was then increased up to 1248 K in a step-by-step ramp-and-dwell manner, as shown in Fig. 1, and maintained at this temperature for 3 h while periodically stirring the ampoules inside the furnace. The gradual increase in furnace temperature, combined with relatively long dwell times, was employed to ensure complete dissolution of solid MnTe and GeTe in the $Sb_2Te_3$ melt. At the end of the synthesis, the samples were first quenched at 1248 K by immediately immersing the ampoules in iced water and then subjected to thermal treatment at 723 K for one month to achieve full homogenization. At the end of the annealing process, the samples in the form of ingots were divided into several pieces and employed for chemical analysis and subsequent physical measurements.

### 2.2. Experimental methods

The surface morphology of the samples was investigated via scanning electron microscopy (SEM) using the JSM-6700F field emission scanning electron microscope (FE-SEM). The elemental analysis of the samples was carried out through energy-dispersive X-ray spectroscopy (EDS) using the QUANTAX-EDS 2.1 system. The elemental compositions derived from the EDS measurements were utilized to determine the experimental chemical formulas and corresponding molar masses of the MGST-124 samples. The structural analysis was performed via powder X-ray diffraction (PXRD) using a Bruker D2 Phaser benchtop diffractometer, equipped with a Cu Kα radiation source (λ = 1.54184 Å) and a LYNXEYE_2 detector operated in 1D mode under Bragg–Brentano geometry. Diffraction data were collected over a 2θ range of 5° to 120°, with a step size of 0.022°, a counting time of 0.139 s per step, and a PSD opening of 5.818°. Using DIFFRAC.EVA software, the obtained diffraction data were processed and analyzed through comparison against reference patterns from COD and PDF-4+ databases. To determine the lattice parameters and the unit cell volume, the obtained PXRD patterns were refined in TOPAS software using fundamental parameters approach based on the reference pattern PDF-04-027-0247 [22]. Chebyshev polynomial was utilized for the refinement by taking into account the following instrumental parameters: the same primary and secondary goniometer radii of 141 mm, the same primary and secondary soller slits of 2.5°, and the divergence slit of 6 mm or 36°. The temperature-dependent phase transformations were identified via differential thermal analysis (DTA) using LINSEIS HDSC PT 1600 coupled with a TC-08 Thermocouple Data Logger. A heating rate of 10 K/min was used, while only the heating curves were utilized for subsequent analysis. The instrument was calibrated using the standard melting points of pure metals (Au, In, Sb, and Sn). DTA measurements were conducted in alumina crucibles under a flowing argon atmosphere with a flow rate of approximately 50 mL/min. The small pieces of the polycrystalline samples cut from the ingots were employed for DC bulk magnetization measurements. Temperature- and field-dependent magnetization measurements were carried out through a commercial superconducting quantum interference device (SQUID) magnetometer (MPMS3, Quantum Design) over the temperature range of 2 K to 300 K and magnetic fields up to ±7 T. The instrument was calibrated with a Pd standard prior to magnetic measurements. The absence of the ferromagnetic impurities was confirmed by pure paramagnetic behavior at room-temperature (300*K*).

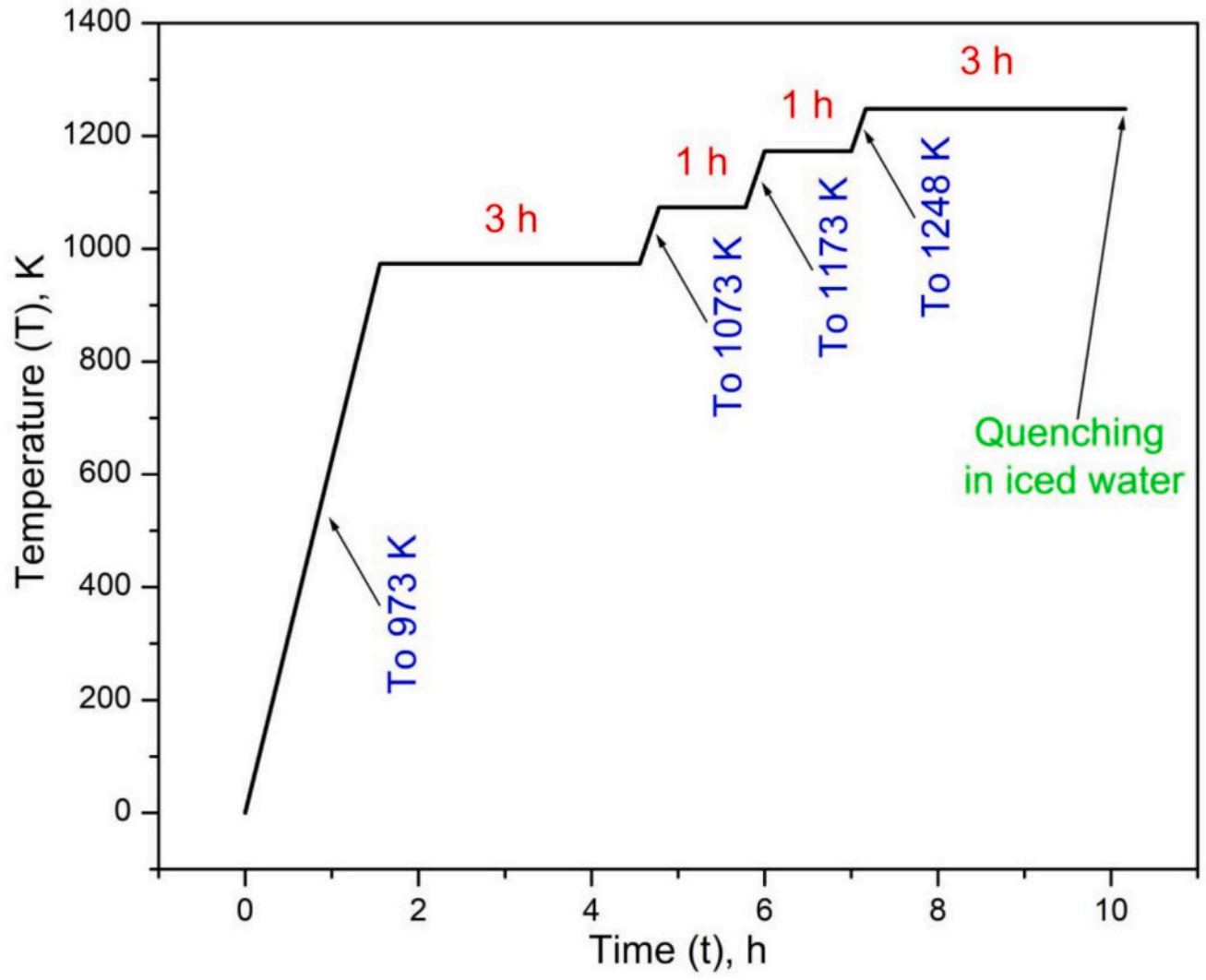

**Fig. 1.** Temperature profile applied during synthesis of the MGST-124 samples. A ramp rate of 7.5 K/min was used. Temperature is given in Kelvin (K), while time is expressed in hours (h).

## 3. Results and discussion

### 3.1. Composition analysis

The EDS spectrum in Fig. S1 shows peaks corresponding to the expected constituent elements Mn, Ge, Sb, and Te, as well as Si and C in one of the $Mn_{1-x}Ge_xSb_2Te_4$ samples with $x = 0.55$ Ge substitution. The Si signal originates from the quartz tube used during synthesis, while the C signal arises from the carbon tape used to mount the sample during EDS measurements. Although Si and C are detected alongside the expected elements, their amounts are negligible and do not affect the structure or properties: therefore, they were excluded from quantification. Table S1 presents the elemental quantification results of the sample with $x = 0.55$ Ge substitution based on the EDS spectra and confirms homogeneity of elemental distribution by showing high uniformity in measurements taken from three different locations from the surface of the sample. The obtained compositions closely match the nominal stoichiometry within experimental error. In addition, Table S2 lists the nominal and experimental atomic percentages of Mn, Ge, Sb, and Te for the MGST-124 samples with error bars representing one standard deviation, indicating minimal differences between nominal and experimental values. The EDS analysis results summarized in Table 1 provide both the nominal and experimentally determined chemical formulas and molar masses of the MGST-124 samples, which are in good agreement with nominal stoichiometry and molar masses of the corresponding $Mn_{1-x}Ge_xSb_2Te_4$ solid solutions, with slight deviations due to minor excesses of Mn, Ge, and Sb and a small deficiency of Te. To determine the chemical formulas and molar masses, the atomic percentages obtained from the EDS spectra were normalized to yield a total of 7 atoms per formula unit. The resulting Ge substitution levels, denoted with $x$, and the experimentally determined molar masses were used in the analysis and comparison of magnetic data.

### 3.2. Structural analysis

The parent ternary $MnSb_2Te_4$ compound belongs to a family of layered transition metal chalcogenides and adopts a tetradymite-like structure, crystallizing in a rhombohedral system with a space group of $R\overline{3}m$ (No. 166), according to a prediction [21] followed by subsequent experimental realizations in polycrystalline [22] and single crystal [31] forms. Structurally, it consists of MnTe bilayers intercalated between quintuple layers of $Sb_2Te_3$, creating a septuple layer (SL) block with atomic sequence of Te-Sb-Te-Mn-Te-Sb-Te, as shown in Fig. 2. These SLs stack along crystallographic c axis and are held together by weak van der Waals (vdW) interactions, whereas interactions within a single SL are dominated by strong covalent bonds. From a crystallographic perspective, the structure appears to be hexagonal when viewed along the ab-plane, but exhibits a rhombohedral symmetry when viewed along crystallographic c axis. Within a single SL, two types of tellurium atoms exist: $Te_1$ atoms occupy the outermost of the SL and participate in interlayer vdW interactions between neighboring SLs, while $Te_2$ atoms are located between Mn and Sb layers and contribute to the intralayer covalent bonding network. Mn atoms occupy the 3a Wyckoff position at the center of the SL, while Sb, $Te_1$, and $Te_2$ atoms are located at the 6c Wyckoff positions.

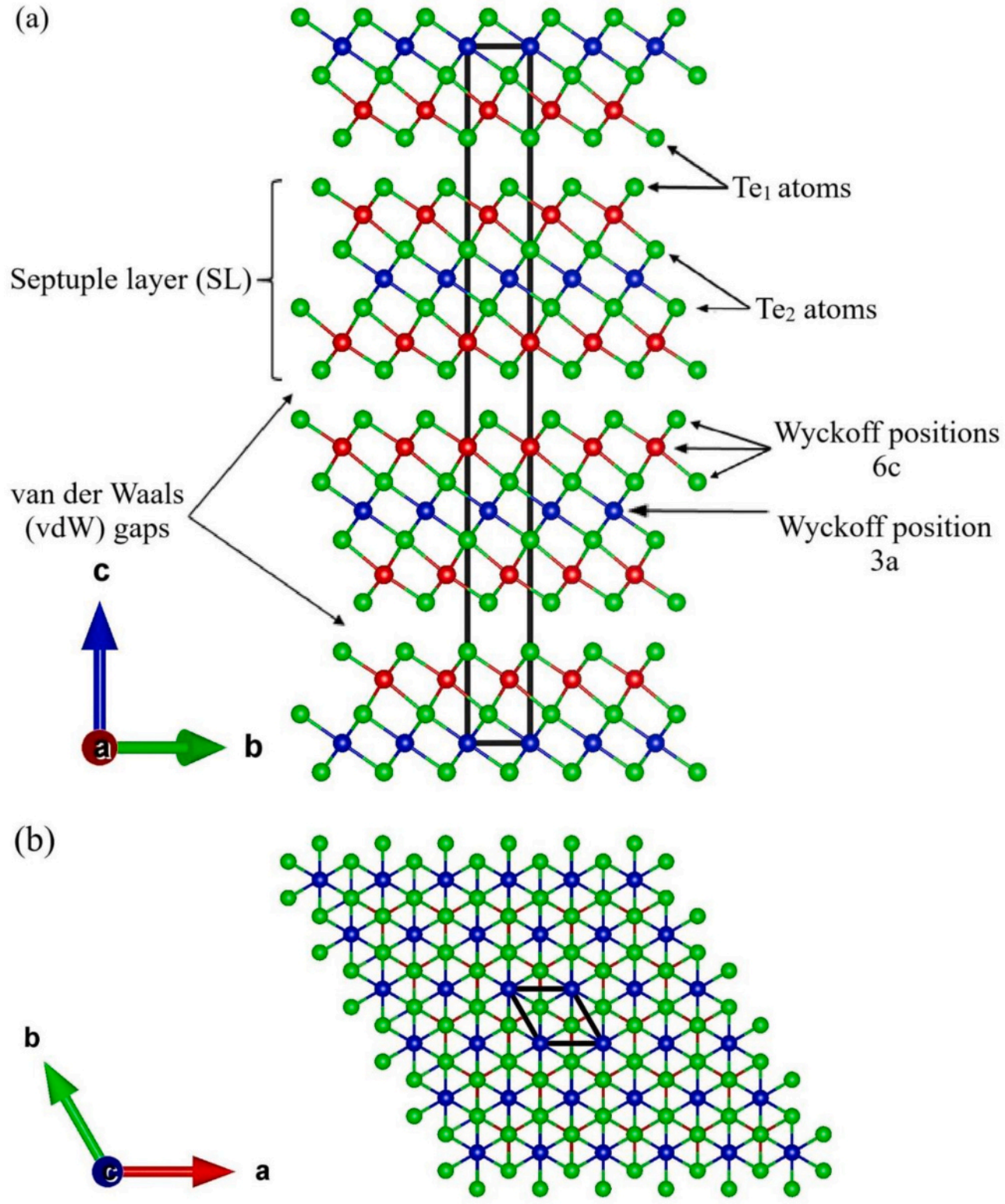

**Fig. 2.** Crystal structure of the parent compound $MnSb_2Te_4$. The unit cell is outlined in dark black. Blue, red, and green spheres represent Mn, Sb, and Te atoms, respectively. The hexagonal or in-plane view is shown in (a), while the rhombohedral or out-of-plane view is indicated in (b). (For interpretation of the references to colour in this figure legend, the reader is referred to the web version of this article.)

**Table 1**
The nominal and experimentally determined chemical formulas and corresponding molar masses of the MGST-124 samples based on the elemental quantification results from the obtained EDS spectra. Sub., nom., and exp. mean substitution, nominal, and experimental, respectively. The molar mass is given in g/mol.

| Sample | Nom. Ge sub. $x$ | Nom. chemical formula | Nom. molar mass in g/mol | Exp. chemical formula | Exp. molar mass in g/mol | Exp. Ge sub. $x$ |
|---|---|---|---|---|---|---|
| 1 | 1 | $Ge_{1.00}Sb_{2.00}Te_{4.00}$ | 826.55 | $Ge_{1.05}Sb_{2.06}Te_{3.89}$ | 823.53 | 1.05 |
| 2 | 0.9 | $Mn_{0.10}Ge_{0.90}Sb_{2.00}Te_{4.00}$ | 824.78 | $Mn_{0.10}Ge_{0.94}Sb_{2.09}Te_{3.86}$ | 821.63 | 0.94 |
| 3 | 0.7 | $Mn_{0.30}Ge_{0.70}Sb_{2.00}Te_{4.00}$ | 821.24 | $Mn_{0.35}Ge_{0.75}Sb_{2.08}Te_{3.82}$ | 814.28 | 0.75 |
| 4 | 0.5 | $Mn_{0.50}Ge_{0.50}Sb_{2.00}Te_{4.00}$ | 817.70 | $Mn_{0.55}Ge_{0.55}Sb_{2.08}Te_{3.82}$ | 811.18 | 0.55 |
| 5 | 0.3 | $Mn_{0.70}Ge_{0.30}Sb_{2.00}Te_{4.00}$ | 814.17 | $Mn_{0.78}Ge_{0.32}Sb_{2.08}Te_{3.82}$ | 806.72 | 0.32 |
| 6 | 0.1 | $Mn_{0.90}Ge_{0.10}Sb_{2.00}Te_{4.00}$ | 810.63 | $Mn_{0.99}Ge_{0.12}Sb_{2.06}Te_{3.84}$ | 803.13 | 0.12 |
| 7 | 0 | $Mn_{1.00}Sb_{2.00}Te_{4.00}$ | 808.86 | $Mn_{1.22}Sb_{1.97}Te_{3.81}$ | 792.99 | 0 |

Due to their identical crystal structures, both crystallizing in $R\bar{3}m$ space group (No. 166) [22,32], pristine $MnSb_2Te_4$ and $GeSb_2Te_4$ samples exhibit qualitatively indistinguishable powder X-ray diffraction (PXRD) patterns, as can be seen from the topmost and bottommost patterns in Fig. 3. The diffraction patterns of the intermediate solid solutions are likewise nearly identical to those of the end-member compounds, as indicated in Fig. 3, indicating the single-phase purity of the substituted compositions. The diffraction peaks observed in the measured samples can be consistently indexed using reference patterns of pristine $MnSb_2Te_4$ available in the PDF-4+ database, such as PDF-04-027-0247 [22], which is depicted in Fig. S2 for the sample with $x = 0.55$ Ge substitution. Furthermore, the observed diffraction peaks are in good agreement with those reported in the literature for polycrystalline and powdered single-crystal samples of pristine $MnSb_2Te_4$ [26,27,33–39]. In addition, a closer examination of the diffraction patterns reveals a gradual shift of the diffraction peaks toward lower 2θ angles with increasing Ge substitution. This shift, resulting from the larger ionic size of $Ge^{2+}$ compared to $Mn^{2+}$, indicates the successful incorporation of Ge atoms into the crystal lattice.

The lattice parameters $a$ and $b$ $(a = b)$ do not exhibit a clear or systematic trend with increasing Ge substitution. This is likely due to the minimal difference between the in-plane lattice parameters of $MnSb_2Te_4$ and $GeSb_2Te_4$, making any variation in $a$ and $b$ too subtle to produce a significant effect. However, the lattice parameter c and also the unit cell volume $V$ increase approximately linearly with Ge substitution, in accordance with Vegard's law, as shown in Fig. 4. This linear expansion along crystallographic c axis, reflecting the larger ionic size of $Ge^{2+}$ compared to $Mn^{2+}$, confirms the successful incorporation of Ge atoms into the crystal structure.

### 3.3. Phase equilibria

Taking into account the previously discussed PXRD results, as well as literature data on phase equilibria in the MnTe–GeTe [40], MnTe–$Sb_2Te_3$ [26], and GeTe–$Sb_2Te_3$ [41] systems, the $T$–$x$ phase diagram of the $MnSb_2Te_4$ – $GeSb_2Te_4$ system was constructed by interpreting the DTA heating curves. As can be seen from Fig. 5, within a wide compositional range (≤ 90 mol% $GeSb_2Te_4$), $\alpha$ solid solutions based on MnTe crystallize first from the melt, followed by the formation of $\gamma$ ($Mn_{1-x}Ge_xSb_2Te_4$) solid solutions through the $L + \alpha \leftrightarrow \gamma$ *(1)* peritectic reaction. While this peritectic reaction occurs as an invariant reaction at 920 K in the MnTe – $Sb_2Te_3$ system [26], in the system under investigation it is monovariant, meaning that it takes place over a temperature interval. This leads to the formation of a three-phase region *($L + \alpha + \gamma$)* on the $T$–$x$ diagram. Since both initial phases – $L$ and $\alpha$ – are entirely consumed in the reaction *(1)*, crystallization ends with the formation of a homogeneous $\gamma$-phase.

In the $GeSb_2Te_4$-rich region of the phase diagram, the crystallization behavior is somewhat different. It is known that the $GeSb_2Te_4$ compound forms via the $L + Ge_2Sb_2Te_5 \leftrightarrow GeSb_2Te_4$ *(2)* peritectic reaction [41]. Therefore, during its crystallization, $Ge_2Sb_2Te_5$ crystallizes first from the melt at 908 K, followed by the crystallization of $GeSb_2Te_4$ at 900 K via reaction (2). This indicates that in the $GeSb_2Te_4$-rich region (≥ 90 mol% $GeSb_2Te_4$) of the system under study, $\delta$-phase based on $Ge_2Sb_2Te_5$ forms first from the melt, followed by the formation of the $\gamma$-phase via reaction (2). Since these processes occur within a very narrow temperature and compositional range, the corresponding phase fields are conventionally outlined with dashed lines.

### 3.4. Magnetic properties

Although pure Mn element is known to be antiferromagnetic with $T_N = 96$ K [42], its interactions with the surrounding crystal structure in the $MnSb_2Te_4$ can be considered as an overall ferrimagnetic ordering [43,44]. The temperature-dependent molar susceptibility data of the MGST-124 samples, indicated in Fig. 6, exhibit a clear bifurcation between the zero-field cooling (ZFC) and field cooling (FC) curves, indicating magnetic ordering competition at low temperatures. However, this bifurcation diminishes with increasing Ge substitution, suggesting the suppression of intralayer ferromagnetic (FM) interactions within the SLs. Furthermore, for the samples with $x = 0.32$ and $x = 0.55$ Ge substitution, an antiferromagnetic-like cusp appears in both the ZFC and FC curves, with the feature being more pronounced in the ZFC curve. This observation highlights the increasing role of intralayer antiferromagnetic (AFM) interactions. The ZFC and FC molar susceptibility curves as functions of temperature, depicted in Figs. S3 and S4, display a systematic decrease in molar susceptibility for the MGST-124 samples with increasing Ge substitution.

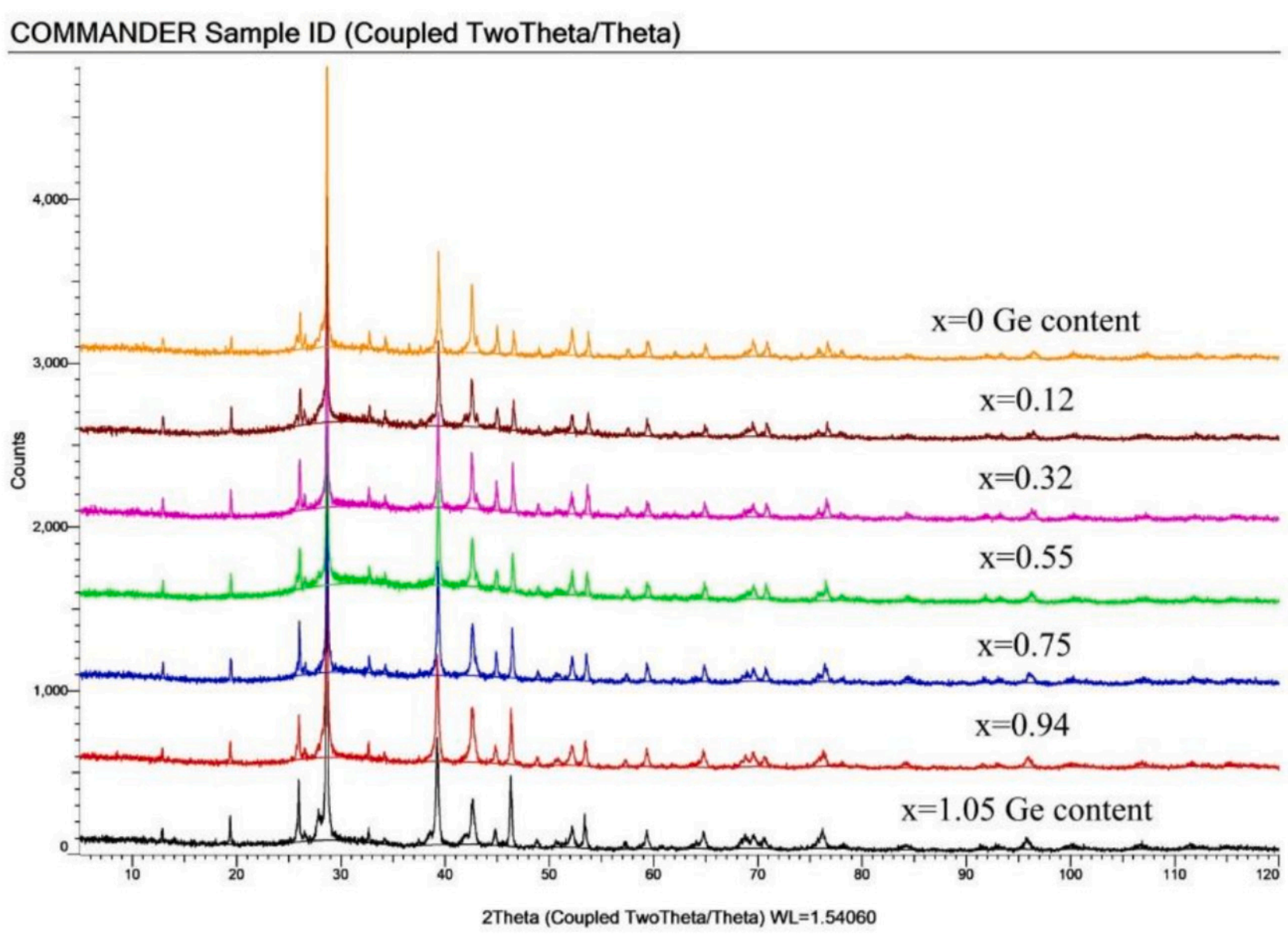

**Fig. 3.** Room-temperature PXRD patterns of the MGST-124 samples with different Ge substitutions ($x$). The intensity is given in arbitrary units (a.u.), while the angle is expressed in 2θ.

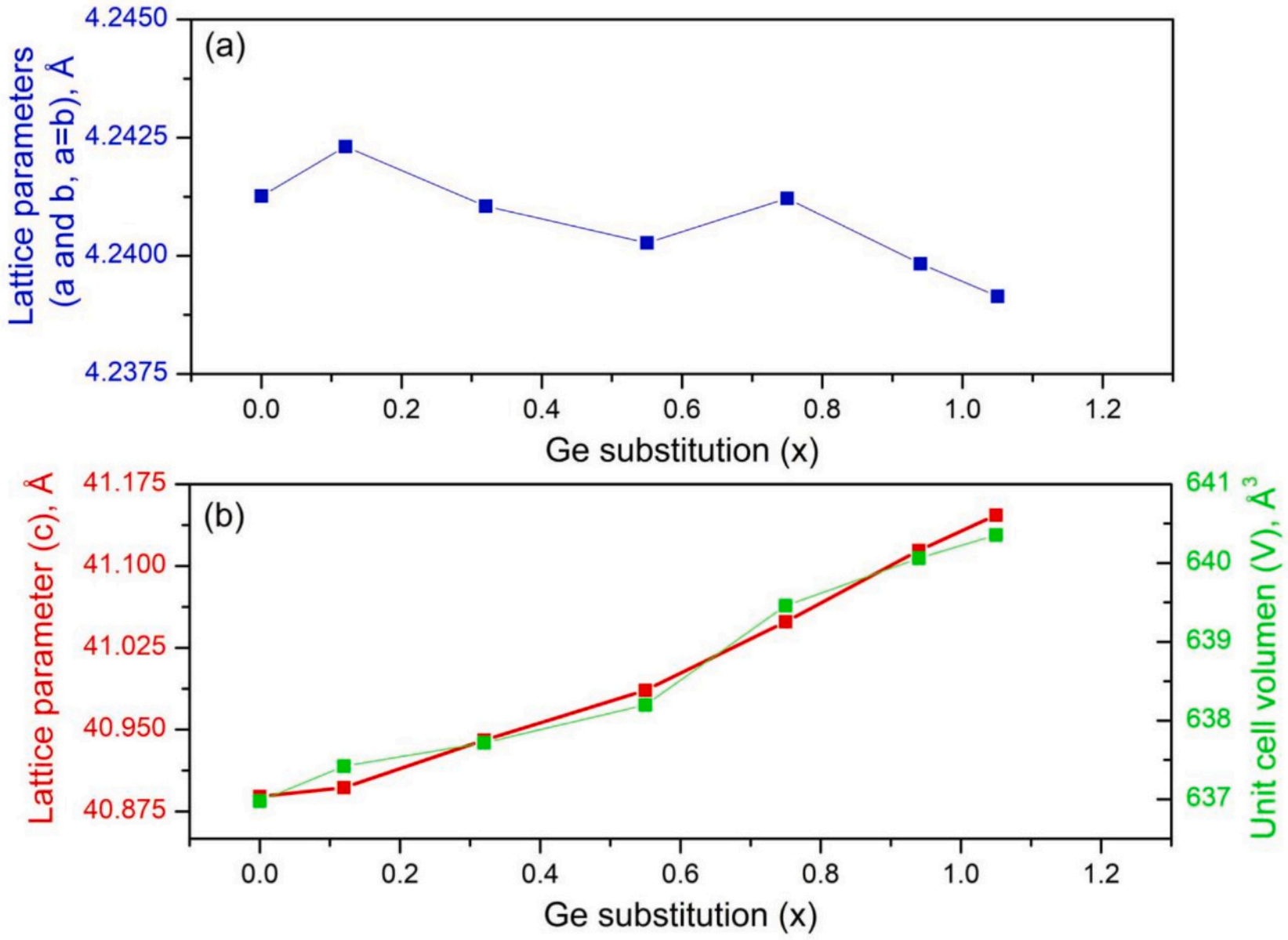

**Fig. 4.** Lattice parameters *a, b, (a* = *b)* and *c*, along with the unit cell volume *V* in the $Mn_{1-x}Ge_xSb_2Te_4$ samples as functions of Ge substitution (*x*). Lattice parameters are given in ångströms (Å), while the unit cell volume is expressed in cubic ångströms ($Å^3$).

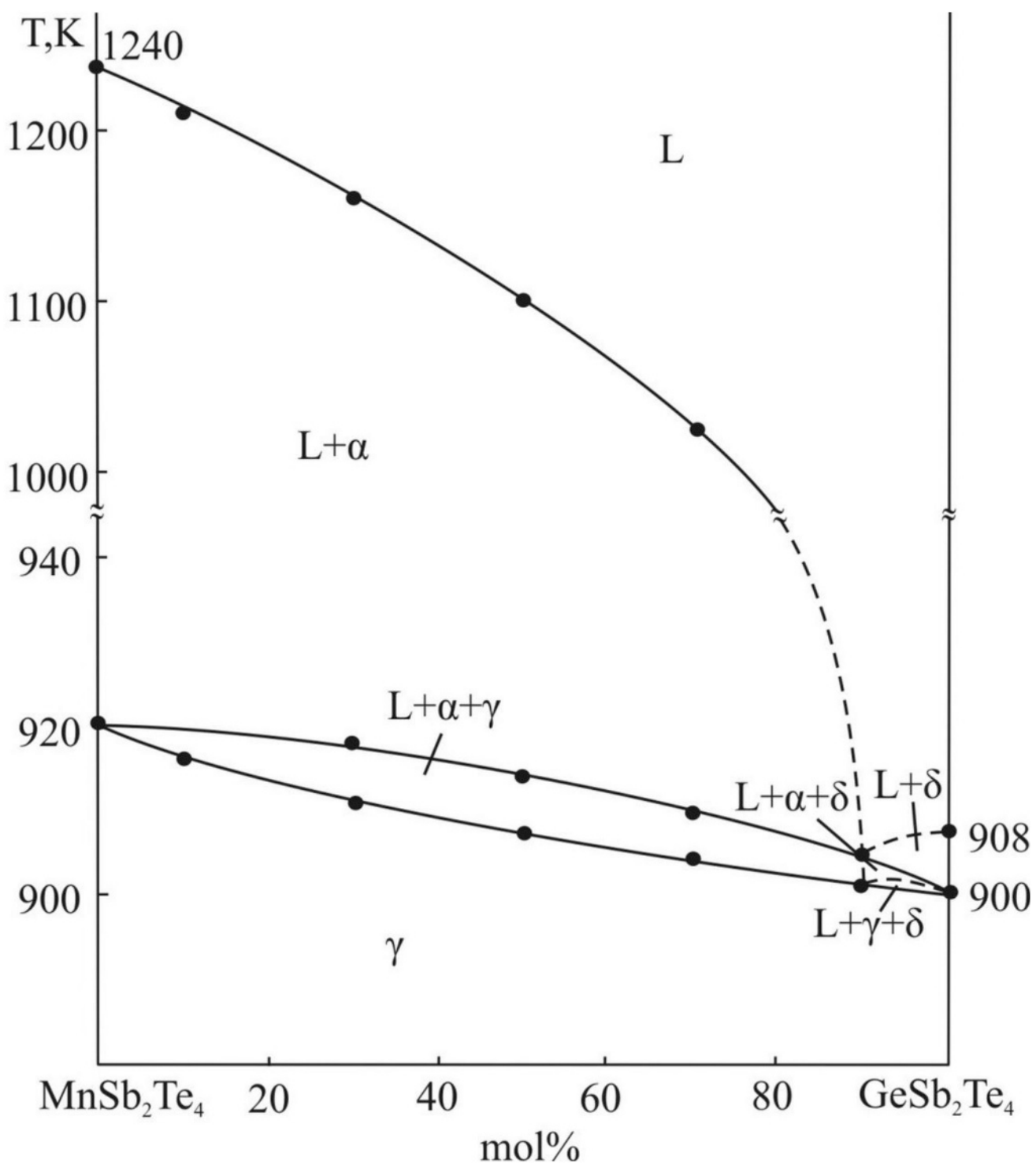

**Fig. 5.** *T*–*x* phase diagram of the $MnSb_2Te_4$ – $GeSb_2Te_4$ system. The temperature is given in Kelvin (K), and the composition is expressed in mol % of $GeSb_2Te_4$. The region labeled *L* corresponds to the liquid phase, while the *α, δ,* and *γ* phases indicate solid solutions based on MnTe, $Ge_2Sb_2Te_5$, and $Mn_{1-x}Ge_xSb_2Te_4$, respectively.

Interestingly, negative magnetization was observed in the ZFC molar susceptibility curves of the pristine $MnSb_2Te_4$ sample and the sample with $x = 0.12$ Ge substitution, though this behavior has never been reported for the former. The observed negative magnetization of the samples that were demagnetized with the ZFC runs underscores the presence of a competition between different magnetic orderings due to existence of magnetic sublattices. Due to Mn—Sb cation intermixing in $MnSb_2Te_4$, Mn and Sb antisite defects occur, in which Mn atoms occupy Sb sites at the 6c Wyckoff position and conversely Sb atoms occupy Mn sites at the 3a Wyckoff position. It had been revealed that the magnetic moment of Mn atoms located at original Mn sites ($Mn_{Mn}$) and that of Mn atoms occupying Sb sites ($Mn_{Sb}$) couple antiferromagnetically, leading to a net positive magnetic moment due to the larger magnetic moments of $Mn_{Mn}$ atoms compared to those of $Mn_{Sb}$ atoms [22]. Achieving full saturation by breaking this AFM coupling between $Mn_{Mn}$ and $Mn_{Sb}$ atoms requires a strong external magnetic field of approximately 60 T to align all spins along the field direction [45].

The strong exchange coupling between $Mn_{Mn}$ and $Mn_{Sb}$ sublattices pins the $Mn_{Mn}$ moments in a direction opposite to the applied magnetic field. Since a magnetic field of 100 Oe applied during the ZFC run lies within the hysteresis loops of the two aforementioned samples with the same coercive field of 200 Oe at 5 K (Fig. 9), the system remains in a minor loop state influenced by its magnetic history. Because the applied magnetic field is insufficient to overcome this pinning, the net magnetization appears negative, reflecting a small but detectable exchange bias effect (Fig. 9), which has also been reported for few-layer $MnSb_2Te_4$ flakes through RMCD measurements [46]. By contrast, during the FC run, a magnetic field of 100 Oe is applied continuously while cooling the samples from room temperature (300 K) through the magnetic ordering temperature down to 2 K. This external magnetic field biases the formation and alignment of magnetic domains, guiding both $Mn_{Mn}$ and $Mn_{Sb}$ sublattices to align more favorably with the magnetic field direction. As a result, the net magnetization remains positive upon warming, even though the magnetic field is still within the hysteresis loop. The first derivative of the molar susceptibility with respect to temperature, $d\chi_n/dT$, reveals a low-temperature magnetic transition from paramagnetic to ferrimagnetic behavior for all the Mn-containing samples, except for the one with $x = 0.94$ Ge substitution Fig. S5. The derivation confirms two distinct magnetic transitions in the samples. The first magnetic transition from paramagnetic to ferrimagnetic, occurring around 24 K, shifts systematically to lower temperatures with increasing Ge substitution, indicating a decrease in the Curie temperature. This ferrimagnetic ordering temperature (~24 K) for the pristine $MnSb_2Te_4$ sample is close to values reported for polycrystalline samples in the literature [22,37,44]. The second metamagnetic transition from ferrimagnetic to ferromagnetic ordering, observed at lower temperatures, displays a dome-like behavior, reaching a maximum for the $x = 0.32$ Ge substitution. The Curie temperatures corresponding to both transitions are indicated in Fig. 7. This second transition may involve gradual freezing or disappearance of $Mn_{Sb}$ magnetic moments, which has been reported for $MnBi_2Te_4$ and its higher homologues [37] (See Fig. 7).

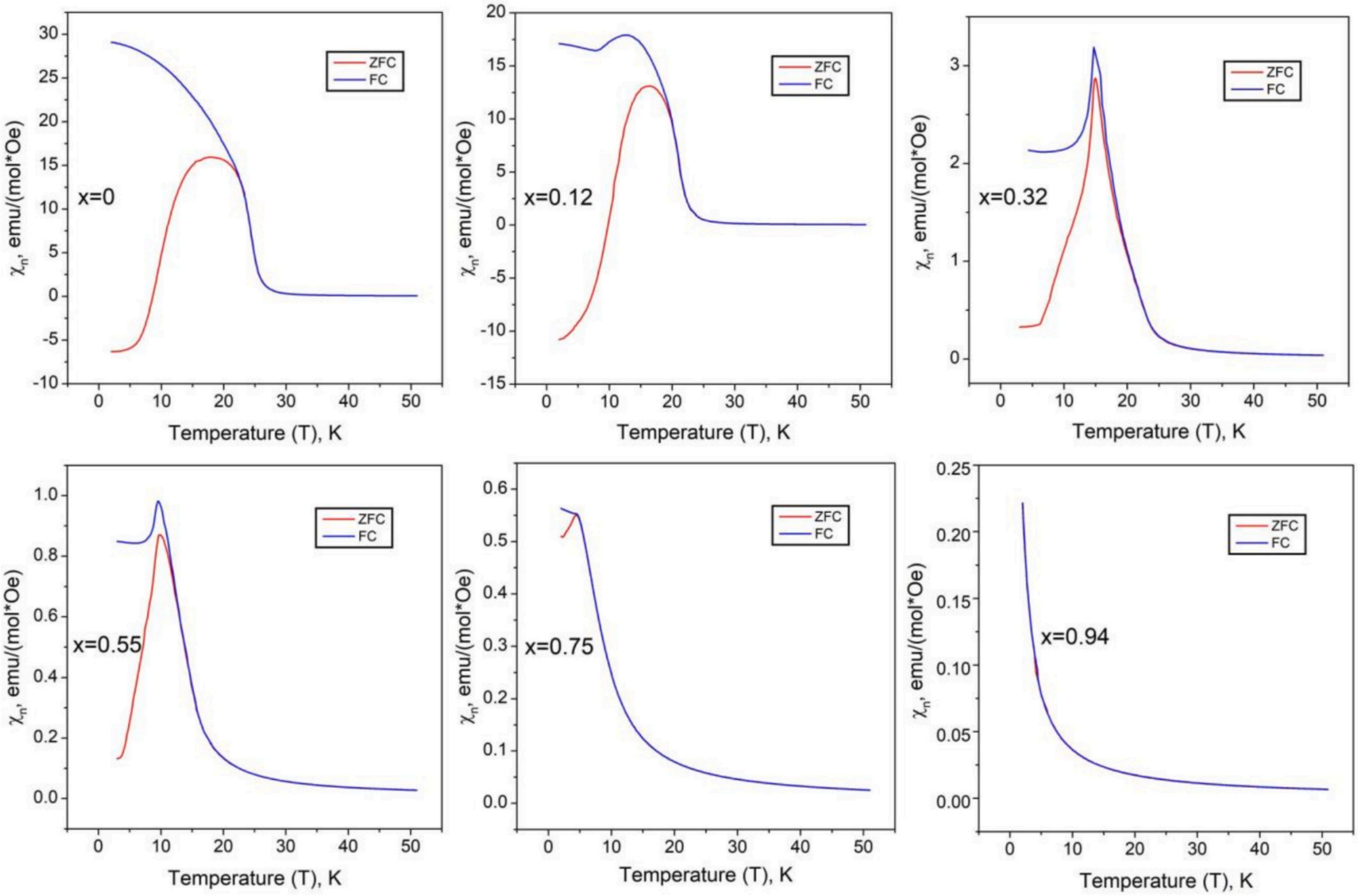

**Fig. 6.** ZFC and FC curves of molar susceptibility as functions of temperature for the MGST-124 samples with different Ge substitutions ($x$). Molar susceptibility is given in emu/(mol·Oe), while temperature is expressed in Kelvin (K).

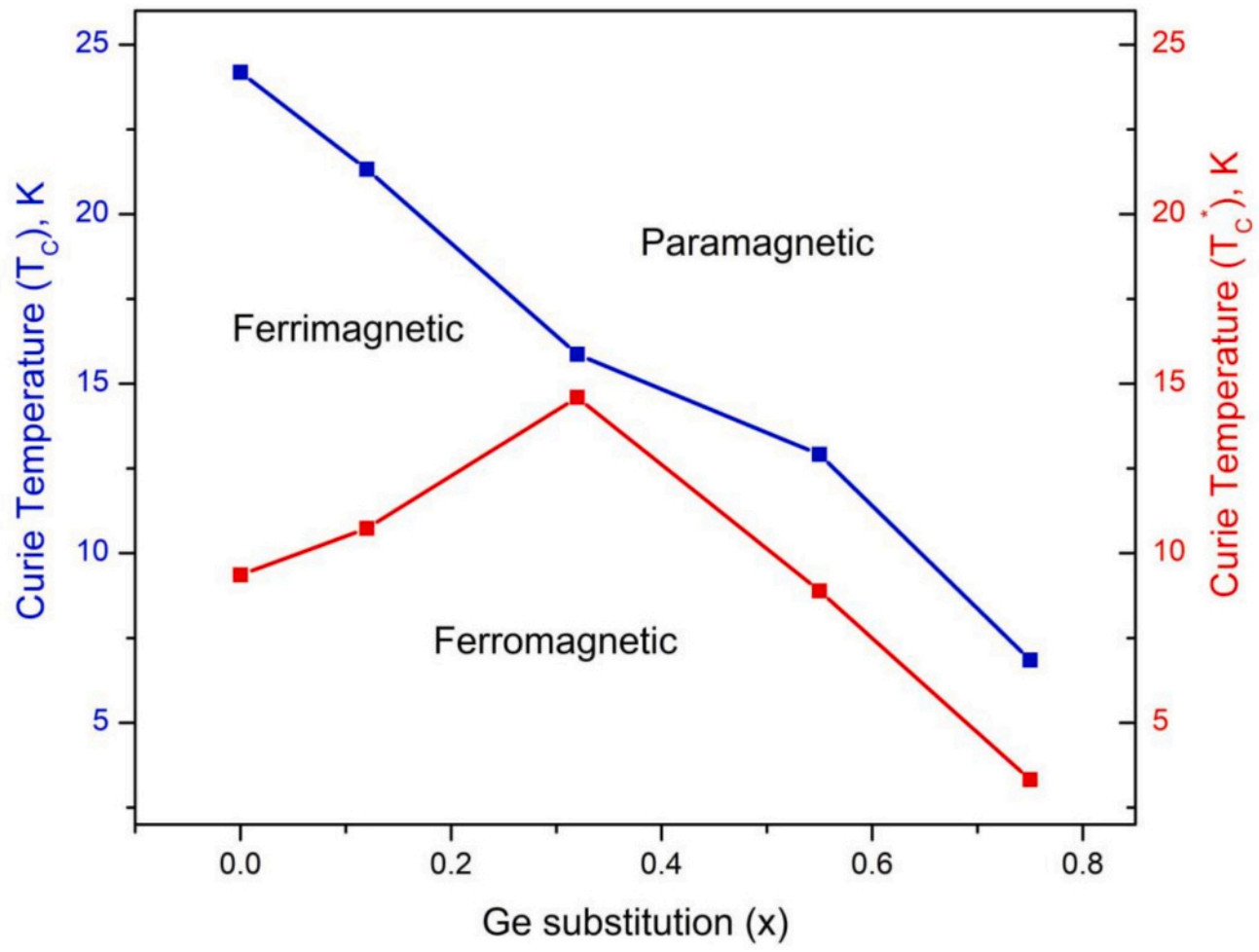

**Fig. 7.** $T_C$-$x$ magnetic phase diagram based on the magnetic data of the MGST-124 samples at 10 mT. Two magnetic transition temperatures (high-temperature $T_C$ and low-temperature $T_C^*$) as functions of Ge substitution ($x$) are given in Kelvin (K).

Both the ZFC and FC inverse molar susceptibility vs temperature data fit well to the linearized equation of the Curie-Weiss law in the high temperature range from 100 K to 300 K, as depicted in Figs. S6 and S7. The inverse molar susceptibility increases with increasing Ge substitution, indicating an enhanced paramagnetic behavior. Notably, the sample with $x = 0.94$ Ge substitution becomes paramagnetic, in contrast to the diamagnetic behavior of $GeSb_2Te_4$. The extracted parameters of Curie constant and Weiss temperature, along with the calculated effective magnetic moment are shown in Figs. S8 and 8 (See Fig. 8) as functions of Ge substitution ($x$). The Curie constant, and consequently the calculated effective magnetic moment, exhibit a systematic evolution and decrease with increasing Ge substitution ($x$). By contrast, the Weiss temperature, which was found to be negative for all the MGST-124 samples, follows a nonlinear dependence on Ge substitution ($x$) with a dome-shape like behavior, reaching a maximum at $x = 0.55$ Ge substitution. The negative Weiss temperature implies dominant antiferromagnetic (AFM) interactions, therefore, the sample with $x = 0.55$ Ge substitution possesses the weakest antiferromagnetic (AFM) interactions among the MGST-124 samples. The extracted Weiss temperature of −18.26 K and the calculated effective moment of 5.42 $\mu_B$ for the pristine $MnSb_2Te_4$ sample are close to values reported for polycrystalline samples in the literature [22]. The discrepancies can be attributed to differences in the degree of Mn—Sb cation intermixing between the samples. Furthermore, the obtained value of 5.42 $\mu_B$ is also close to the theoretical value of 5.92 $\mu_B$ expected for $Mn^{2+}$ ions in the high spin state.

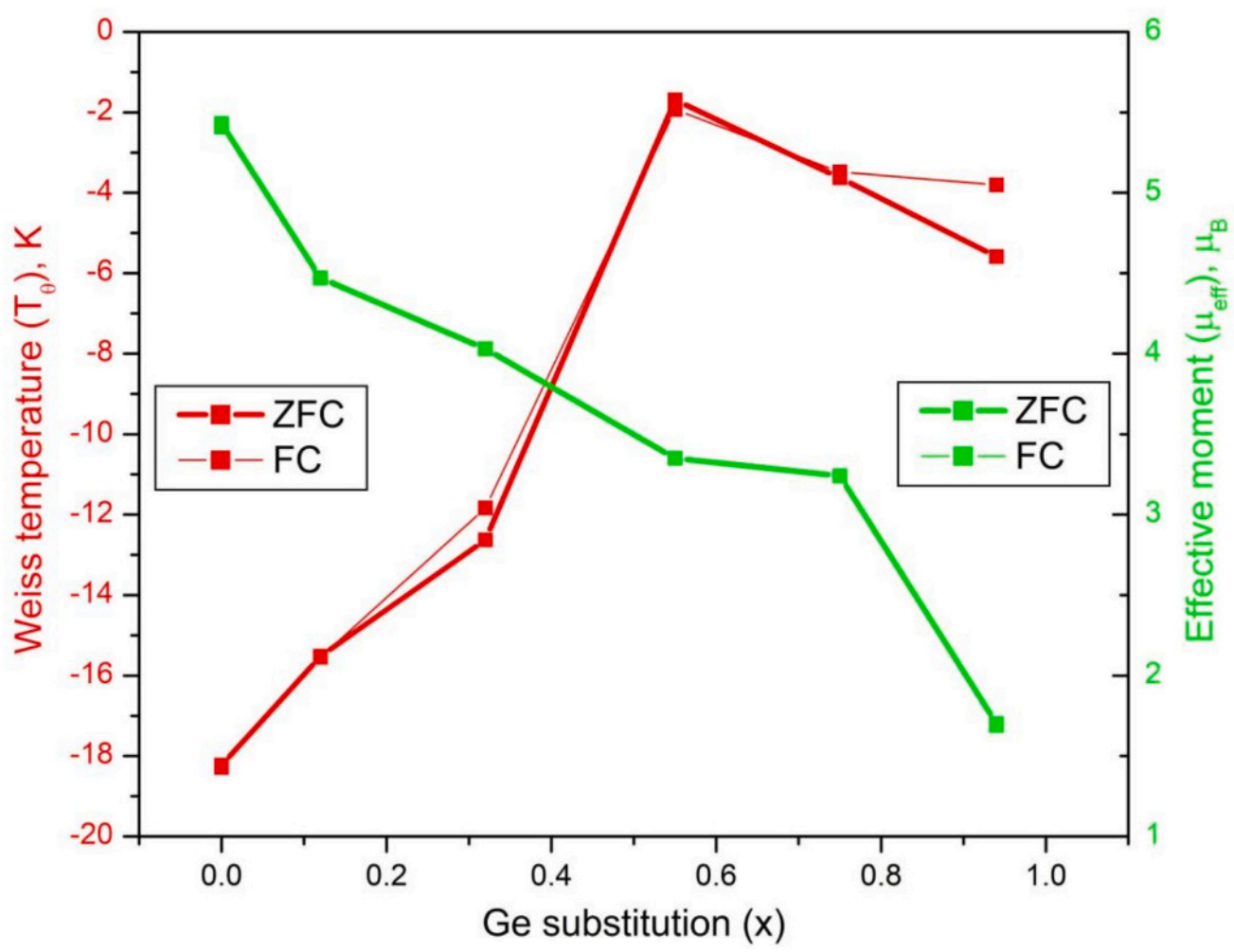

**Fig. 8.** The extracted Weiss temperature and the calculated effective moment as functions of Ge substitution ($x$) for both ZFC and FC runs. The former is given in Kelvin (K), while the latter is expressed in Bohr magnetons ($\mu_B$).

Isothermal magnetization, loops (*M(H)*) measured at 5 K in the zero-field ZFC mode reveal a systematic evolution of magnetic behavior in $Mn_{1-x}Ge_xSb_2Te_4$ solid solutions with increasing Ge substitution, as indicated in Fig. 9. As the Ge content increases, the *M(H)* loops evolve from ferromagnetic-like to ferrimagnetic-like behavior, accompanied by a gradual decrease in the saturation magnetic moment as depicted in Figs. S9 and S10 and a progressive reduction of hysteresis followed by its eventual disappearance. Both the pristine $MnSb_2Te_4$ and the composition with the lowest Ge substitution of $x = 0.12$ show small hysteresis loops with different remanent magnetizations of 0.46 $\mu_B$/f.u. and 0.32 $\mu_B$/f.u., respectively. Though the pristine compound exhibits a higher remanent magnetization than the Ge-substituted sample, both possess the same coercive field of 200 Oe. Furthermore, the pristine $MnSb_2Te_4$ displays a low-field saturation magnetic moment of 1.99 $\mu_B$/f.u, which is closer to values reported for polycrystalline samples [22], single crystals [24,35,39,47–55], and thin films [20,56] in the literature.

Due to Mn-Sb cation intermixing, the observed saturation magnetic moment is significantly lower than the theoretical value of 5.0 $\mu_B$ expected for $Mn^{2+}$ ions in the high spin state. Mn-Sb antisite defects lead to a compensation effect due to antiferromagnetic coupling between larger magnetic moments of $Mn_{Mn}$ atoms and smaller magnetic moments of $Mn_{Sb}$ atoms [22], thereby reducing the overall saturation magnetic moment. Achieving a full saturation requires a strong external magnetic field of approximately 60 T to align all spins along the magnetic field direction [45]. The $x = 0.12$ Ge-substituted sample shows a slightly reduced saturation magnetic moment of 1.77 $\mu_B$/f.u., which is 12% lower than that of the pristine compound.

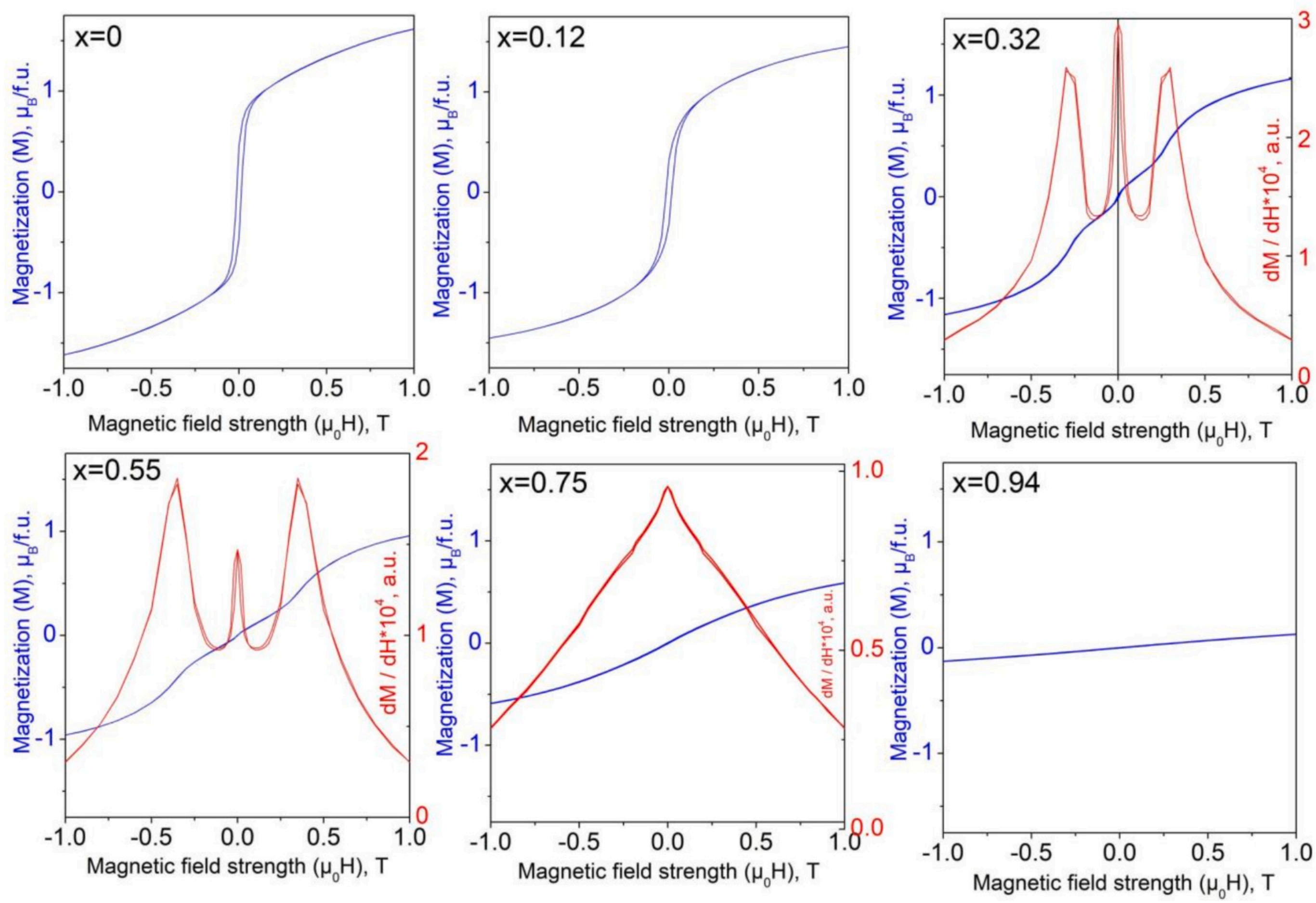

**Fig. 9.** ZFC isothermal field-dependent magnetization loops at 5 K for the MGST-124 samples with different Ge substitutions ($x$). The left y axis shows the magnetization ($M$) in Bohr magnetons per formula unit ($\mu_B$/f.u.), while the right y axis displays the first derivative of magnetization ($M$) with respect to magnetic field ($H$), ($dM/dH$), to better visualize spin-flop metamagnetic transitions. The derivatives are given in arbitrary units (a.u.), while magnetic field strength is expressed in tesla (T).

Increasing the Ge substitution from $x = 0.12$ to $x = 0.32$ gives rise to spin-flop-like metamagnetic behavior, characterized by two distinct transitions at magnetic fields of 1.2 kOe and 3.5 kOe. These transitions indicate the emergence and strengthening of field-induced ferromagnetic interactions at higher magnetic fields. The spin-flop features are more distinctly resolved by plotting the first derivative of magnetization ($M$) with respect to the magnetic field ($H$), $dM/dH$, which highlights the abrupt slope changes associated with the metamagnetic transitions, as depicted in Fig. 9. A similar two-step behavior is observed in the sample with $x = 0.55$ Ge substitution, featuring spin-flop transitions at 1.4 kOe and 3.0 kOe. Interestingly, compared to the $x = 0.32$ Ge-substituted composition, the first spin-flop in the sample with $x = 0.55$ Ge substitution occurs at a slightly higher field, while the second one appears at a lower field, suggesting a complex and non-monotonic evolution of magnetic exchange interactions with increasing Ge substitution.

The emergence of spin-flop-like behavior can be attributed to magnetic dilution caused by Ge substitution. As non-magnetic Ge content increases the average distance between Mn atoms increases as well, which weakens the intralayer FM coupling between Mn atoms. When this coupling becomes sufficiently weak, canted antiferromagnetic (CAFM) [57,58] or purely AFM interactions may develop between Mn atoms. Under an external magnetic field, these CAFM- or AFM-coupled spins undergo reorientation to align with the field, resulting in a metamagnetic transition. In the $x = 0.12$ Ge-substituted sample, the intralayer FM coupling remains strong enough to suppress AFM correlations, and thus no spin-flop transition is observed. Conversely, in the $x = 0.75$ Ge-substituted sample, the intralayer FM coupling is so strongly suppressed that CAFM and AFM interactions dominate. As a result, no field-induced spin-flop transition is observed. Though no spin-flop behavior appears, the sample with $x = 0.75$ Ge substitution still exhibits ferrimagnetism with a saturation magnetic moment of 1.01 $\mu_B$/f.u., which is lower than those of $x = 0.55$ and $x = 0.32$ Ge-substituted compositions by 23% and 31% respectively. The sample with the highest Ge substitution of $x = 0.94$ was found to be paramagnetic with no low-temperature magnetic transition and a much reduced saturation magnetic moment of 0.33 $\mu_B$/f.u.

## 4. Conclusion

In this work, we investigated phase equilibria in the $MnSb_2Te_4$ – $GeSb_2Te_4$ system and magnetic properties of Ge-substituted $MnSb_2Te_4$ solid solutions. SEM-EDS and PXRD analysis showed homogeneity and single-phase purity for all the prepared solid solutions, while the Rietveld refinement identified an overall lattice expansion with increases in the lattice parameter c and the unit cell volume V in accordance with Vegard's law, though no clear or systematic trend was observed for the lattice parameters a and b. Both composition and structural analysis together with the Rietveld refinement confirm the successful incorporation of Ge into Mn sites at the crystal structure of $MnSb_2Te_4$ in the full substitution range. The phase diagram of the $MnSb_2Te_4$ – $GeSb_2Te_4$ system was constructed, once again confirming that homogenous and single-phase $Mn_{1-x}Ge_xSb_2Te_4$ solid solutions exist in the full substitution range and all melt by peritectic reaction. Magnetic measurements revealed that increasing Ge substitution progressively enhances paramagnetism and intralayer AFM interactions in the samples, while the magnetic transition persists up to $x = 0.75$. Two distinct magnetic transitions – high-temperature PM to FIM and low-temperature FIM to FM – were observed, the latter of which originates from a gradual freezing or disappearance of magnetic moments of Mn atoms occupying Sb sites. The first magnetic transition temperature followed a systematic trend with increasing Ge substitution, whereas the second magnetic transition temperature exhibited a dome-shape like dependence, reaching a maximum at $x = 0.32$. The FM ordering temperature was found to increase for $x \leq 0.32$ and decrease for larger substitution levels. Taking into account large sensitivity to antisite disorder, one can assume that Ge alters the balance of Mn-Sb defects. This can enable new, shorter Mn-Te-Mn exchange paths till optimum substitution level. The exact mechanism determination, however, requires a separate study taking into account other possible contributions. The negative magnetization appearing for the pristine $MnSb_2Te_4$ and $x = 0.12$ samples is attributed to pinning of magnetic moments of Mn atoms occupying original Mn sites, accompanied with a detectable exchange bias effect. The effective magnetic moment and the saturation magnetization exhibited a systematic decrease with increasing Ge substitution. Two spin-flop transitions, indicating existence of competing magnetic orderings, were observed for the $x = 0.32$ and $x = 0.55$ samples, where ferrimagnetic and ferromagnetic tendencies are finely balanced. These results pave the way for future studies on such samples in single crystals form, as well as other related compounds with interesting magnetic properties [59], which are essential for unraveling the interplay between magnetism and topology, ultimately advancing both fundamental understanding and potential applications.

## CRediT authorship contribution statement

**Fuad Safarov:** Writing – original draft, Investigation, Data curation. **Dunya Babanly:** Writing – review & editing, Supervision, Funding acquisition. **Jerome Robert:** Software, Methodology. **Elnur Orujlu:** Supervision, Conceptualization. **Elvin Ahmadov:** Software, Methodology. **Martin Bowen:** Writing – review & editing. **Bohdan Kundys:** Writing – original draft, Supervision, Resources.

## Declaration of competing interest

The authors declare that they have no known competing financial interests or personal relationships that could have appeared to influence the work reported in this paper.

## Acknowledgements

We gratefully acknowledge cofunding from French Embassy in Azerbaijan and French-Azerbaijani University for joint supervision PhD program, and the research grant from Azerbaijan State Oil and Industry University (the grant order number: 3-29-50/3-1-4367/2025). We also thank Nicolas Beyer and Fabien Chevrier for technical assistance, Marc Lenertz and Tom Ferte for PXRD and SEM-EDS measurements/analysis, respectively.

## Appendix A. Supplementary data

Supplementary data to this article can be found online at https://doi.org/10.1016/j.jmmm.2026.173969.

## Data availability

Data will be made available on request.

## References

[1] M.Z. Hasan, C.L. Kane, Colloquium: Topological insulators, Rev. Mod. Phys. 82 (2010) 3045–3067, https://doi.org/10.1103/RevModPhys.82.3045.

[2] X.-L. Qi, S.-C. Zhang, Topological insulators and superconductors, Rev. Mod. Phys. 83 (2011) 1057–1110, https://doi.org/10.1103/RevModPhys.83.1057.

[3] J.E. Moore, The birth of topological insulators, Nature 464 (2010) 194–198, https://doi.org/10.1038/nature08916.

[4] D. Pesin, A.H. MacDonald, Spintronics and pseudospintronics in graphene and topological insulators, Nat. Mater. 11 (2012) 409–416, https://doi.org/10.1038/nmat3305.

[5] C.-Z. Chang, J. Zhang, X. Feng, J. Shen, Z. Zhang, M. Guo, K. Li, Y. Ou, P. Wei, L.-L. Wang, Z.-Q. Ji, Y. Feng, S. Ji, X. Chen, J. Jia, X. Dai, Z. Fang, S.-C. Zhang, K. He, Y. Wang, L. Lu, X.-C. Ma, Q.-K. Xue, Experimental observation of the quantum anomalous hall effect in a magnetic topological insulator, Science 340 (2013) 167–170, https://doi.org/10.1126/science.1234414.

[6] Y. Tokura, K. Yasuda, A. Tsukazaki, Magnetic topological insulators, Nat. Rev. Phys. 1 (2019) 126–143, https://doi.org/10.1038/s42254-018-0011-5.

[7] X.-L. Qi, T.L. Hughes, S.-C. Zhang, Topological field theory of time-reversal invariant insulators, Phys. Rev. B 78 (2008) 195424, https://doi.org/10.1103/PhysRevB.78.195424.

[8] R. Yu, W. Zhang, H.-J. Zhang, S.-C. Zhang, X. Dai, Z. Fang, Quantized anomalous hall effect in magnetic topological insulators, Science 329 (2010) 61–64, https://doi.org/10.1126/science.1187485.

[9] Y.L. Chen, J.-H. Chu, J.G. Analytis, Z.K. Liu, K. Igarashi, H.-H. Kuo, X.L. Qi, S. K. Mo, R.G. Moore, D.H. Lu, M. Hashimoto, T. Sasagawa, S.C. Zhang, I.R. Fisher, Z. Hussain, Z.X. Shen, Massive Dirac fermion on the surface of a magnetically doped topological insulator, Science 329 (2010) 659–662, https://doi.org/10.1126/science.1189924.

[10] W. Luo, X.-L. Qi, Massive Dirac surface states in topological insulator/magnetic insulator heterostructures, Phys. Rev. B 87 (2013) 085431, https://doi.org/10.1103/PhysRevB.87.085431.

[11] P. Li, J. Yu, J. Xu, L. Zhang, K. Huang, Topological-magnetic proximity effect in $Sb_2Te_3/CrI_3$ heterostructures, Phys. B Condens. Matter 573 (2019) 77–80, https://doi.org/10.1016/j.physb.2019.08.009.

[12] M. Mogi, M. Kawamura, R. Yoshimi, A. Tsukazaki, Y. Kozuka, N. Shirakawa, K. S. Takahashi, M. Kawasaki, Y. Tokura, A magnetic heterostructure of topological insulators as a candidate for an axion insulator, Nat. Mater. 16 (2017) 516–521, https://doi.org/10.1038/nmat4855.

[13] D. Xiao, J. Jiang, J.-H. Shin, W. Wang, F. Wang, Y.-F. Zhao, C. Liu, W. Wu, M.H. W. Chan, N. Samarth, C.-Z. Chang, Realization of the Axion insulator state in quantum anomalous hall Sandwich Heterostructures, Phys. Rev. Lett. 120 (2018) 056801, https://doi.org/10.1103/PhysRevLett.120.056801.

[14] K. Nomura, N. Nagaosa, Surface-quantized anomalous hall current and the magnetoelectric effect in magnetically disordered topological insulators, Phys. Rev. Lett. 106 (2011) 166802, https://doi.org/10.1103/PhysRevLett.106.166802.

[15] J. Wang, B. Lian, X.-L. Qi, S.-C. Zhang, Quantized topological magnetoelectric effect of the zero-plateau quantum anomalous hall state, Phys. Rev. B 92 (2015) 081107, https://doi.org/10.1103/PhysRevB.92.081107.

[16] M. Mogi, M. Kawamura, A. Tsukazaki, R. Yoshimi, K.S. Takahashi, M. Kawasaki, Y. Tokura, Tailoring tricolor structure of magnetic topological insulator for robust axion insulator, Sci. Adv. 3 (2017) eaao1669, https://doi.org/10.1126/sciadv.aao1669.

[17] M.M. Otrokov, I.I. Klimovskikh, H. Bentmann, D. Estyunin, A. Zeugner, Z.S. Aliev, S. Gaß, A.U.B. Wolter, A.V. Koroleva, A.M. Shikin, M. Blanco-Rey, M. Hoffmann, I. P. Rusinov, A.Y. Vyazovskaya, S.V. Eremeev, Y.M. Koroteev, V.M. Kuznetsov, F. Freyse, J. Sánchez-Barriga, I.R. Amiraslanov, M.B. Babanly, N.T. Mamedov, N. A. Abdullayev, V.N. Zverev, A. Alfonsov, V. Kataev, B. Büchner, E.F. Schwier, S. Kumar, A. Kimura, L. Petaccia, G. Di Santo, R.C. Vidal, S. Schatz, K. Kißner, M. Ünzelmann, C.H. Min, S. Moser, T.R.F. Peixoto, F. Reinert, A. Ernst, P. M. Echenique, A. Isaeva, E.V. Chulkov, Prediction and observation of an antiferromagnetic topological insulator, Nature 576 (2019) 416–422, https://doi.org/10.1038/s41586-019-1840-9.

[18] Y. Deng, Y. Yu, M.Z. Shi, Z. Guo, Z. Xu, J. Wang, X.H. Chen, Y. Zhang, Quantum anomalous hall effect in intrinsic magnetic topological insulator $MnBi_2Te_4$, Science 367 (2020) 895–900, https://doi.org/10.1126/science.aax8156.

[19] C. Liu, Y. Wang, H. Li, Y. Wu, Y. Li, J. Li, K. He, Y. Xu, J. Zhang, Y. Wang, Robust axion insulator and Chern insulator phases in a two-dimensional antiferromagnetic topological insulator, Nat. Mater. 19 (2020) 522–527, https://doi.org/10.1038/s41563-019-0573-3.

[20] S. Wimmer, J. Sánchez-Barriga, P. Küppers, A. Ney, E. Schierle, F. Freyse, O. Caha, J. Michalicka, M. Liebmann, D. Primetzhofer, M. Hoffmann, A. Ernst, M. M. Otrokov, G. Bihlmayer, E. Weschke, B. Lake, E.V. Chulkov, M. Morgenstern, G. Bauer, G. Springholz, O. Rader, Mn-rich $MnSb_2Te_4$: a topological insulator with magnetic gap closing at high curie temperatures of 45-50 K, Adv. Mater. 33 (2021) 2102935, https://doi.org/10.1002/adma.202102935.

[21] S.V. Eremeev, M.M. Otrokov, E.V. Chulkov, Competing rhombohedral and monoclinic crystal structures in $MnPn_2Te_4$: an ab-initio study, J. Alloys Compd. 709 (2017) 172–178, https://doi.org/10.1016/j.jallcom.2017.03.121.

[22] T. Murakami, Y. Nambu, T. Koretsune, G. Xiangyu, T. Yamamoto, C.M. Brown, H. Kageyama, Realization of interlayer ferromagnetic interaction in $MnSb_2Te_4$ toward the magnetic Weyl semimetal state, Phys. Rev. B 100 (2019) 195103, https://doi.org/10.1103/PhysRevB.100.195103.

[23] J. Xiong, Y.-H. Peng, J.-Y. Lin, Y.-J. Cen, X.-B. Yang, Y.-J. Zhao, High concentration intrinsic defects in $MnSb_2Te_4$, Materials 16 (2023) 5496, https://doi.org/10.3390/ma16155496.

[24] Y. Liu, L.-L. Wang, Q. Zheng, Z. Huang, X. Wang, M. Chi, Y. Wu, B. C. Chakoumakos, M.A. McGuire, B.C. Sales, W. Wu, J. Yan, Site mixing for engineering magnetic topological insulators, Phys. Rev. X 11 (2021) 021033, https://doi.org/10.1103/PhysRevX.11.021033.

[25] M. Nurmamat, K. Okamoto, S. Zhu, T.V. Menshchikova, I.P. Rusinov, V. O. Korostelev, K. Miyamoto, T. Okuda, T. Miyashita, X. Wang, Y. Ishida, K. Sumida, E.F. Schwier, M. Ye, Z.S. Aliev, M.B. Babanly, I.R. Amiraslanov, E.V. Chulkov, K. A. Kokh, O.E. Tereshchenko, K. Shimada, S. Shin, A. Kimura, Topologically nontrivial phase-change compound $GeSb_2Te_4$, ACS Nano 14 (2020) 9059–9065, https://doi.org/10.1021/acsnano.0c04145.

[26] E. Orujlu, Z. Aliev, I. Amiraslanov, M. Babanly, Phase equilibria of the MnTe-$Sb_2Te_3$ system and synthesis of novel ternary layered compound – $MnSb_4Te_7$, Phys. Chem. Solid State 22 (2021) 39–44, https://doi.org/10.15330/pcss.22.1.39-44.

[27] Y. Luo, J. Wang, J. Yang, D. Mao, J. Cui, B. Jia, X. Liu, K. Nielsch, X. Xu, J. He, Microstructural iterative reconstruction toward excellent thermoelectric performance in MnTe, Energy Environ. Sci. 16 (2023) 3743–3752, https://doi.org/10.1039/D3EE01902K.

[28] M.N. Mamontov, Optimization of thermodynamic data on liquidus of Mn-Te phase diagram in associated-solution model; Termodinamicheskoe soglasovanie dannykh po likvidusu fazovoj diagrammy sistemy Mn-Te s pomoshch'yh modeli assotsiirovannykh rastvorov, Neorg. Mater. 32 (1996). https://www.osti.gov/etdeweb/biblio/518952.

[29] D. Bletskan, Phase equilibrium in the system AIV-BVI - part II: systems germanium-chalcogen, J. Ovonic Res. 1 (2005) 53–60.

[30] S. Solé, C. Schmetterer, K.W. Richter, A revision of the Sb-Te binary phase diagram and crystal structure of the modulated γ-phase field, J. Phase Equilib. Diffus. 43 (2022) 648–659, https://doi.org/10.1007/s11669-022-00958-5.

[31] J.-Q. Yan, S. Okamoto, M.A. McGuire, A.F. May, R.J. McQueeney, B.C. Sales, Evolution of structural, magnetic, and transport properties in $MnBi_{2-x}Sb_xTe_4$, Phys. Rev. B 100 (2019) 104409, https://doi.org/10.1103/PhysRevB.100.104409.

[32] K.A. Agaev, A.G. Talybov, Electron-diffraction analysis of structure of $GeSb_2Te_4$, Sov. Phys. Crystallogr. USSR 11 (1966) 400.

[33] S.B. Izzatli, E.N. Orujlu, Y.I. Jafarov, M.B. Babanly, Phase relations and tunable solid solutions in the pseudoternary system MnTe– $Sb_2Te_3$–$Bi_2Te_3$, J. Chem. Eng. Data 70 (2025) 1759–1768, https://doi.org/10.1021/acs.jced.4c00711.

[34] X. Xu, D. Mao, M.O. Liedke, M. Butterling, J. Cui, J. Wang, Y. Luo, Z. Ge, E. Hirschmann, A. Wagner, J. He, K. Nielsch, R. He, Advancing carriers mobility in $MnSb_2Te_4$ thermoelectrics via tailored textures and vacancy modification, Adv. Energy Mater. 15 (2025) 2500838, https://doi.org/10.1002/aenm.202500838.

[35] H. Li, Y. Li, Y. Lian, W. Xie, L. Chen, J. Zhang, Y. Wu, S. Fan, Glassy magnetic ground state in layered compound $MnSb_2Te_4$, Sci. China-Mater. 65 (2022) 477–485, https://doi.org/10.1007/s40843-021-1738-9.

[36] T. Yu, T. Zhang, X. Qu, N. Qi, D. Yuan, X. Su, G. Tan, X. Tang, Z. Chen, Significant enhancement in the thermoelectric performance of the $MnSb_2Te_4$ topological insulator through vacancy regulation and lattice-softening strategies, ACS Appl. Mater. Interfaces 17 (2025) 14229–14242, https://doi.org/10.1021/acsami.4c21679.

[37] M. Sahoo, I.J. Onuorah, L.C. Folkers, E.V. Chulkov, M.M. Otrokov, Z.S. Aliev, I. R. Amiraslanov, A.U.B. Wolter, B. Büchner, L.T. Corredor, C. Wang, Z. Salman, A. Isaeva, R. De Renzi, G. Allodi, Ubiquitous order-disorder transition in the Mn antisite sublattice of the $(MnBi_2Te_4)(Bi_2Te_3)_n$, 2024, https://doi.org/10.48550/arXiv.2402.06340.

[38] E.B.P. Journal, E. Orujlu, Phase equilibria in the $SnBi_2Te_4$-$MnBi_2Te_4$ system and characterization of the $Sn_{1-x}Mn_xBi_2Te_4$ solid solutions, Phys. Chem. Solid State 21 (2020) 113–116, https://doi.org/10.15330/pcss.21.1.113-116.

[39] H. Li, Y. Li, Y.-K. Lian, W. Xie, L. Chen, J. Zhang, Y. Wu, S. Fan, Spin glass state in layered compound $MnSb_2Te_4$, 2021, https://doi.org/10.48550/arXiv.2104.00898.

[40] W.D. Johnston, D.E. Sestrich, The MnTe-GeTe phase diagram, J. Inorg. Nucl. Chem. 19 (1961) 229–236, https://doi.org/10.1016/0022-1902(61)80111-5.

[41] Azerbaijan State Oil and Industry University, E.N. Orujlu, Reinvestigation of the GeTe- Sb2Te3 pseudobinary phase diagram, Azerbaijan Chem. J. 0 (2025) 95–103, https://doi.org/10.32737/0005-2531-2025-2-95-103.

[42] J.M.D. Coey, Magnetism and Magnetic Materials, Cambridge University Press, 2010.

[43] T.A. Webb, A.N. Tamanna, X. Ding, N. Verma, J. Xu, L. Krusin-Elbaum, C.R. Dean, D.N. Basov, A.N. Pasupathy, Tunable magnetic domains in ferrimagnetic $MnSb_2Te_4$, Nano Lett. 24 (2024) 4393–4399, https://doi.org/10.1021/acs.nanolett.3c05058.

[44] M. Sahoo, Z. Salman, G. Allodi, A. Isaeva, L. Folkers, A. Wolter, B. Büchner, R. De Renzi, Impact of Mn-Pn intermixing on magnetic properties of an intrinsic magnetic topological insulator: the μSR perspective, J. Phys. Conf. Ser. 2462 (2023) 012040, https://doi.org/10.1088/1742-6596/2462/1/012040.

[45] Y. Lai, L. Ke, J. Yan, R.D. McDonald, R.J. McQueeney, Defect-driven ferrimagnetism and hidden magnetization in $MnBi_2Te_4$, Phys. Rev. B 103 (2021) 184429, https://doi.org/10.1103/PhysRevB.103.184429.

[46] Z. Zang, M. Xi, S. Tian, R. Guzman, T. Wang, W. Zhou, H. Lei, Y. Huang, Y. Ye, Exchange Bias effects in ferromagnetic $MnSb_2Te_4$ down to a monolayer, ACS Appl. Electron. Mater. 4 (2022) 3256–3262, https://doi.org/10.1021/acsaelm.2c00568.

[47] X. Hu, X. He, Z. Guo, T. Kamiya, J. Wu, Antisite-defects control of magnetic properties in $MnSb_2Te_4$, ACS Nano 18 (2024) 738–749, https://doi.org/10.1021/acsnano.3c09064.

[48] X. Yang, J. Pan, X. He, D. Chu, Critical behavior, magnetic phase diagram, and magnetic entropy change of $MnSb_2Te_4$, Phys. Rev. B 109 (2024) 094408, https://doi.org/10.1103/PhysRevB.109.094408.

[49] M. Xi, F. Chen, C. Gong, S. Tian, Q. Yin, T. Qian, H. Lei, Relationship between antisite defects, magnetism, and band topology in $MnSb_2Te_4$ crystals with $T_C \approx 40$ K, J. Phys. Chem. Lett. 13 (2022) 10897–10904, https://doi.org/10.1021/acs.jpclett.2c02775.

[50] D.Y. Yan, M. Yang, P.B. Song, Y.T. Song, C.X. Wang, C.J. Yi, Y.G. Shi, Site mixing induced ferrimagnetism and anomalous transport properties of the Weyl semimetal candidate $MnSb_2Te_4$, Phys. Rev. B 103 (2021) 224412, https://doi.org/10.1103/PhysRevB.103.224412.

[51] S. Huan, D. Wang, H. Su, H. Wang, X. Wang, N. Yu, Z. Zou, H. Zhang, Y. Guo, Magnetism-induced ideal Weyl state in bulk van der Waals crystal $MnSb_2Te_4$, Appl. Phys. Lett. 118 (2021) 192105, https://doi.org/10.1063/5.0047438.

[52] W. Ge, P.M. Sass, J. Yan, S.H. Lee, Z. Mao, W. Wu, Direct evidence of ferromagnetism in $MnSb_2Te_4$, Phys. Rev. B 103 (2021) 134403, https://doi.org/10.1103/PhysRevB.103.134403.

[53] Z. Zang, Y. Zhu, M. Xi, S. Tian, T. Wang, P. Gu, Y. Peng, S. Yang, X. Xu, Y. Li, B. Han, L. Liu, Y. Wang, P. Gao, J. Yang, H. Lei, Y. Huang, Y. Ye, Layer-number-dependent antiferromagnetic and ferromagnetic behavior in $MnSb_2Te_4$, Phys. Rev. Lett. 128 (2022) 017201, https://doi.org/10.1103/PhysRevLett.128.017201.

[54] J.-Q. Yan, Z. Huang, W. Wu, A.F. May, Vapor transport growth of $MnBi_2Te_4$ and related compounds, J. Alloys Compd. 906 (2022) 164327, https://doi.org/10.1016/j.jallcom.2022.164327.

[55] A.N. Tamanna, A. Lakra, X. Ding, E. Buzi, K. Park, K. Sobczak, H. Deng, G. Sharma, S. Tewari, L. Krusin-Elbaum, Transport chirality generated by a tunable tilt of Weyl nodes in a van der Waals topological magnet, Nat. Commun. 15 (2024) 9830, https://doi.org/10.1038/s41467-024-53319-w.

[56] S. Wimmer, J. Sánchez-Barriga, P. Küppers, A. Ney, E. Schierle, F. Freyse, O. Caha, J. Michalicka, M. Liebmann, D. Primetzhofer, M. Hoffmann, A. Ernst, M. Otrokov, G. Bihlmayer, E. Weschke, B. Lake, E. Chulkov, M. Morgenstern, G. Bauer, O. Rader, Ferromagnetic MnSb2Te4: a topological insulator with magnetic gap closing at high Curie temperatures of 45–50 K, 2020.

[57] S.-K. Bac, K. Koller, F. Lux, J. Wang, L. Riney, K. Borisiak, W. Powers, M. Zhukovskyi, T. Orlova, M. Dobrowolska, J.K. Furdyna, N.R. Dilley, L. P. Rokhinson, Y. Mokrousov, R.J. McQueeney, O. Heinonen, X. Liu, B.A. Assaf, Topological response of the anomalous hall effect in $MnBi_2Te_4$ due to magnetic canting, Npj Quantum Mater. 7 (2022) 46, https://doi.org/10.1038/s41535-022-00455-5.

[58] D. Lujan, J. Choe, M. Rodriguez-Vega, Z. Ye, A. Leonardo, T.N. Nunley, L.-J. Chang, S.-F. Lee, J. Yan, G.A. Fiete, R. He, X. Li, Magnons and magnetic fluctuations in atomically thin $MnBi_2Te_4$, Nat. Commun. 13 (2022) 2527, https://doi.org/10.1038/s41467-022-29996-w.

[59] O.J. Amin, A. Dal Din, E. Golias, Y. Niu, A. Zakharov, S.C. Fromage, C.J.B. Fields, S. L. Heywood, R.B. Cousins, F. Maccherozzi, J. Krempaský, J.H. Dil, D. Kriegner, B. Kiraly, R.P. Campion, A.W. Rushforth, K.W. Edmonds, S.S. Dhesi, L. Šmejkal, T. Jungwirth, P. Wadley, Nanoscale imaging and control of altermagnetism in MnTe, Nature 636 (2024) 348–353, https://doi.org/10.1038/s41586-024-08234-x.